\documentclass[twocolumn]{aastex631}

\newcommand{\valer}[2]{#1^{+#2}_{-#2}}

\newcommand{\SRC}[6]{#1 ($P_{orb}=#2\,\text{min}$, $T_1=\valer{#3}{#4}\,\text{K}$, $T_2=\valer{#5}{#6}\,\text{K}$)}
\newcommand{\SRCS}[4]{#1 ($P_{orb}=#2\,\text{min}$, $T_1=\valer{#3}{#4}\,\text{K}$)}
\newcommand{\src}[6]{#1 ($#2\,\text{min}$, $\valer{#3}{#4}\,\text{K}$, $\valer{#5}{#6}\,\text{K}$)}
\newcommand{\srcs}[4]{#1 ($#2\,\text{min}$, $\valer{#3}{#4}\,\text{K}$)}

\begin{document}

\title{The Temperature versus Orbital Period relation of AM~CVns: Insights from their Donors}

\author[0000-0002-9209-2787]{Colin W. Macrie}
\affiliation{Rutgers University Department of Physics and Astronomy,
136 Frelinghuysen Road,
Piscataway, NJ 08854, USA}
\affiliation{Dept. of Physics and Astronomy, University of Texas Rio Grande Valley, Brownsville, TX 78520, USA}

\author[0000-0002-9396-7215]{Liliana Rivera Sandoval}
\affiliation{Dept. of Physics and Astronomy, University of Texas Rio Grande Valley, Brownsville, TX 78520, USA}

\author[0000-0002-6447-3603]{Yuri Cavecchi}
\affiliation{Departament de Fisica, EEBE, Universitat Politecnica de Catalunya, Av. Eduard Maristany 16, E-08019 Barcelona, Spain}

\author[0000-0001-9195-7390]{Tin Long Sunny Wong}
\affiliation{Department of Physics, University of California, Santa Barbara, CA 93106, USA}

\author[0000-0003-4436-831X]{Manuel Pichardo Marcano}
\affiliation{Department of Life and Physical Sciences, Fisk University, 1000 17th Avenue N., Nashville, TN 37208, USA}
\affiliation{Department of Physics and Astronomy, Vanderbilt University, 6301 Stevenson Center Lane, Nashville, TN 37235, USA}

\begin{abstract}

We studied the spectral energy distribution (SED) of 22 known AM~CVns with orbital periods ($P_{orb}$) larger than 35~min using multiwavelength public photometric data to estimate the effective temperature of the accreting white dwarf. We find an infrared (IR) excess in all systems when compared to a single blackbody, both when the disk should be extended and when it should be truncated by the accretor's magnetic field. This suggests a dominant contribution from the donor to the IR flux. When fitting two blackbodies, the temperature of the hot component decreases with $P_{orb}$, as expected by evolutionary models. Temperatures for systems with $35<P_{orb}<45~\text{min}$ are consistent with models. Systems with $P_{orb}\gtrsim45~\text{min}$ have higher temperatures than expected. The second blackbody temperature  does not correlate with $P_{orb}$. 

\end{abstract}


\section{Introduction} \label{sec:intro}

AM~CVns are binaries harboring accreting white dwarfs (WDs) with $P_{orb}<70$~min, lacking Hydrogen but Helium abundant. The donor is expected to be a He-WD or a semidegenerate He-rich star. Systems with $P_{orb}>20~\text{min}$ show outbursts and superoutbursts similar to dwarf novae in WDs accreting from H-rich stars, with the outburst frequency decreasing with $P_{orb}$. For AM~CVns with $P_{orb}>35~\text{min}$ the outbursts have recurrence times from several years up to multiple decades \citep[$P_{orb}>50~\text{min}$,][]{2015Levitan}.

When $P_{orb}>35~\text{min}$, the disk contribution should be substantially less than in persistent systems or systems with frequent outbursts and one could model the SED with a blackbody to have an estimate of the accretor's effective temperature.

Little is known about the donors in AM~CVns, mainly due to their faintness and the paucity of known systems. 
Donor tentative detections have been reported by \cite{2020Green,2021RS,2022vR}, where an IR excess was identified.

\section{Methodology} \label{sec:obs}

We studied 22 AM~CVns with $P_{orb}>35~\text{min}$. We expect minimal contamination by outbursts to our SEDs because for $35<P_{orb}<45$~min the outbursts are not frequent and the high-state lasts less than a few weeks \citep{2019Canizzo}, while at $P_{orb}>50$~min the disks are mostly stable and cold, {\color{black}hence they display high-state very rarely. We keep both groups to gauge the disk contribution, as systems with $35<P_{orb}<45$~min are expected to have a larger optical/IR flux.} We included Gaia14aae, $P_{orb}=49.7$~min, which \cite{2024Maccarone} have identified as a magnetic system. 

For each system we obtained VizieR photometric data\footnote{http://vizier.cds.unistra.fr/vizier/sed/} using an approximately $5\arcsec$ search radius, optimized via visual inspection in Aladin, to reduce contamination from nearby stars. The data included UV (GALEX, Swift-UVOT, except for systems SDSSJ1730+5545, SDSSJ0129+3842, ZTF18acnnabo, and ZTFJ0003+1404), optical (e.g. Gaia, SDSS, Pan-STARRS) and IR (WISE). We manually discarded bad photometric points.

SEDs were dereddened using Gaia distances \citep{2021BJ} and 3D-Dust Maps \citep{2019Green}, and fit  with one and two blackbodies to estimate the accretor effective temperature and account for additional contributions (e.g. the donor star).

\section{Results} \label{sec:res}

\begin{figure}
    \centering
    \includegraphics[scale=.65]{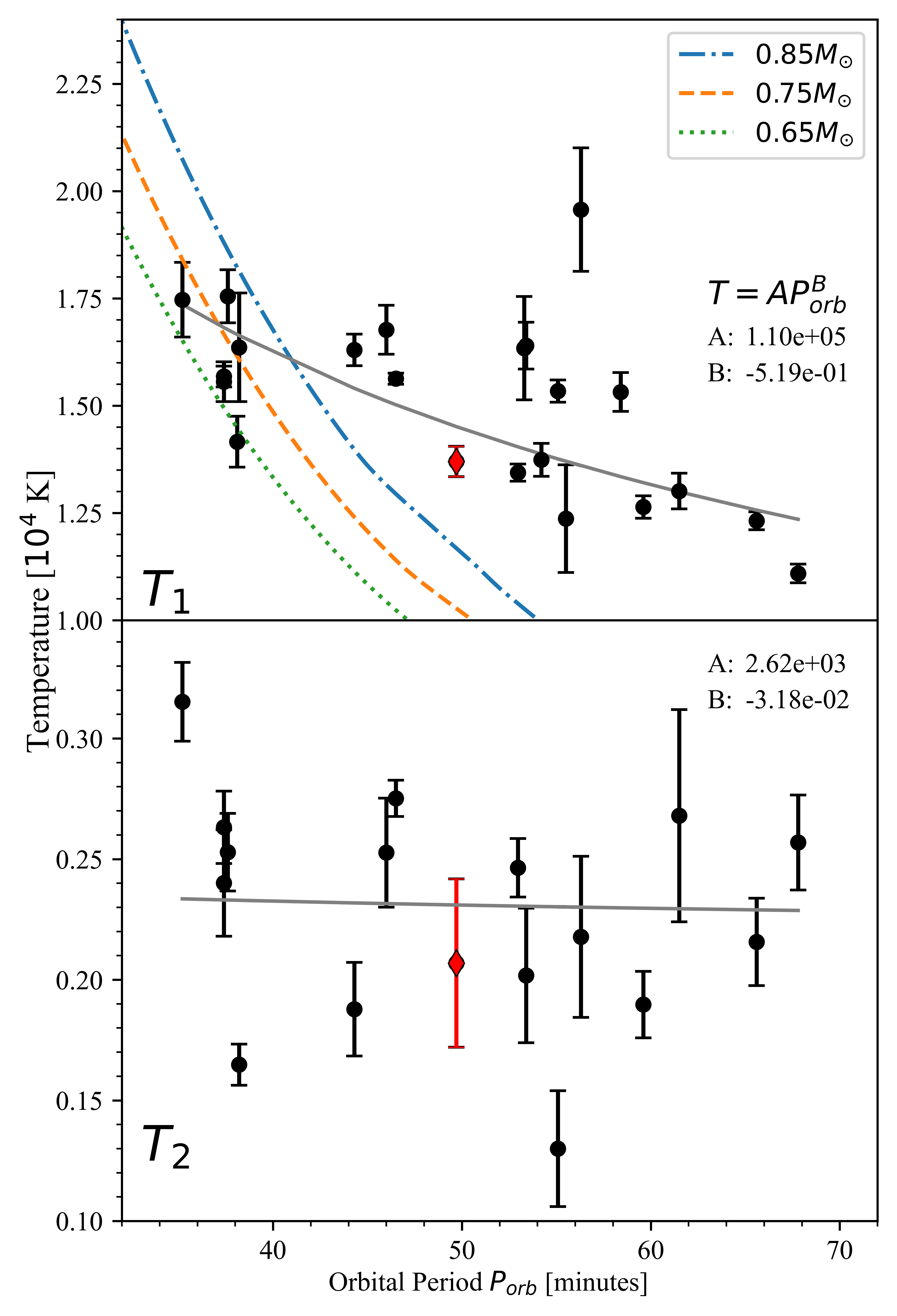}
    \caption{Temperature vs $P_{orb}$
      for the hot ($T_1$, top) and the cold ($T_2$, bottom) components of AM~CVns when fitting a double blackbody to the SEDs, {\color{black}except five systems (see text)}. The magnetic system Gaia14aae is indicated in red.
      We plot models by \cite{2021Wong} for $0.65~M_\odot- 0.85~M_\odot$WDs. The best fits to the $T_1$ and $T_2$ relations are shown with solid grey lines. {\color{black} Results for all the 22 systems analyzed are available in Appendix \ref{sec:Appendix}.}
    }
    \label{fig:fig1}
\end{figure}

Figure \ref{fig:fig1} shows the temperature of the hot ($T_1$) and cold ($T_2$) components vs $P_{orb}$. 
\SRCS{SDSSJ1721+2733}{38.1}{14160}{590}, \srcs{SDSSJ0807+4852}{53}{16340}{1200},
\srcs{SDSSJ1642+1934}{54.2}{13740}{380}, 
\srcs{ZTFJ0003+1404}{55.5}{12370}{1250} and 
\srcs{ASASSN-21au}{58.4}{15320}{450}, which lack IR data,
were well fit with a single blackbody. The fit of $T_1$ is mostly driven by the UV data, and hence is dominated by the accretor. The cold component fit is driven by the IR data. 
We fit both $T-P_{orb}$ relations with the powerlaw model $A P_{orb}^B$.

All systems with available IR data showed an excess when fitting with one blackbody, including the magnetic system Gaia14aae. The binary SDSSJ0804+1616 ($P_{orb}=44.5~\text{min}$, not
in Fig.\ref{fig:fig1}), magnetic, and possibly diskless \citep{2024Maccarone}, also shows IR excess. 

When considering the fit of two blackbodies, all $T_2$ are in the range $1300-3200~\text{K}$, with an average value of $2300~\text{K}$. $T_2$ is similar for both $35<P_{orb}<45~\text{min}$ and $P_{orb}>45~\text{min}$ systems. For Gaia14aae we obtained $T_2=\valer{2070}{350}\,\text{K}$, consistent with the non-magnetic AM~CVns, and $T_1=\valer{1370}{350}\,\text{K}$, close to the spectroscopic value of $\valer{12900}{200}\,\rm{K}$ \citep{2015Campbell}. Contrary to the anticorrelation observed between $T_1$ and $P_{orb}$, for $T_2$ we did not observe any relation with $P_{orb}$. 

Comparing to one blackbody fits reported by \cite{2022vR}, we obtained $T_1$ values consistent within errors for \SRC{ZTFJ2252-0519}{37.4}{15560}{460}{2630}{150} and ZTFJ0003+1404, and $15\%$ higher for \src{ZTFJ0220+2141}{53.4}{16400}{550}{2020}{280} and \src{ZTFJ1637+4917}{61.5}{13010}{420}{2680}{440}. According to these authors, the latter 2 systems have emission dominated by the WD and basically no contribution from the hotspot and disk at optical wavelengths.

\section{Discussion} \label{sec:disc}

We compared the accretor's temperature vs period relation to evolutionary models by \citet{2021Wong}. Systems with $35<P_{orb}<45~\text{min}$ are consistent with models for WDs in the range $0.65-0.85M_{\odot}$, while $P_{orb}\gtrsim45~\text{min}$ systems are hotter than expected {\color{black} by models \citep{2021Wong}, but consistent with results by \cite{2022vR}.}
The higher $T_1$ may be attributed to a greater than expected donor entropy or additional orbital angular momentum loss \citep[e.g. by magnetic braking,][]{2023Belloni}, leading to an increased mass-transfer rate and faster orbital evolution, and consequently a higher accretor temperature at a given $P_{orb}$. This also implies a larger donor mass than expected. These phenomena could help explain the superoutbursts observed in long period AM~CVns \citep{2020RS,2021RS,2021Wongnote,2022RS}, which are attributed to the donor's enhanced mass-transfer.

The IR excess present regardless of whether the disk is hot, cold, extended or magnetically truncated, suggests the emission is dominated by the donor. The higher than expected and nearly constant $T_2$ {\color{black}indicates the donor might be strongly irradiated} (perhaps with non-fully-negligible disk or hotspot contribution). {\color{black}Detailed IR spectroscopic studies and modeling are necessary to
confirm this interpretation.}

Our results show that, without spectra, SEDs are a useful tool to estimate AM CVn temperatures of {\color{black}systems with $P_{orb}>35$ min, where contamination
from high-state is low. Including data in UV and IR
will yield the best constraints.}\\

While  carrying  out this project, we became aware that Courreges et al. were conducting a similar study.

\begin{acknowledgments}
This project was supported by the UTRGV-REU program. We thank student A. Mendoza for early help.
\end{acknowledgments}

\appendix
\section{Figure Data}\label{sec:Appendix}

The following table is available as a machine readable file in the published version on the Research Notes of the American Astronomical Society. 

\begin{table}[h]
    \centering
    \begin{tabular}{c c c c c c}
\hline
Source name & Period & T1 & error T1 &  T2 & error T2\\
             & (min) & (K) & (K) & (K) & (K)\\
\hline
SDSS J1730+5545 & $35.2^a$  & 17470 & 870 &  3152 & 160\\
SDSS J1240-0159 & $37.4^a$  & 15680 & 240 &  2400 & 220\\
ZTFJ2252-0519 &  $37.4^b$  & 15560 & 460 &  2630 & 150\\
SDSS J0129+3842 & $37.6^a$  & 17550 & 620 &  2530 & 160 \\
SDSS J1721+2733 & $38.1^a$ & 14160 & 590 &       &  \\
ZTF18acnnabo  & $38.2^c$  & 16360 & 1270 & 1650 & 85\\
SDSS J1525+3600 & $44.3^a$  & 16300 & 370 &  1880 & 195	\\
SDSS J1411+4812 & $46.0^d$  &   16770 & 570 &  2525 & 225	\\
GP Com  &  $46.6^a$  & 15630 & 130 &  2750 & 75\\
Gaia14aae  &  $49.7^a$  & 13700 & 355 &  2070 & 350	\\
SDSS J1208+3550 & $53.0^a$  &  13440 & 200 &  2465 & 120\\
SDSS J0807+4852 & $53.3^e$  & 16340 & 1200 & &\\
ZTFJ0220+2141 &  $53.3^b$  & 16400 & 550 &  2020 & 280\\
SDSS J1642+1934 & $54.2^a$  & 13740 & 385 & &\\
SRGeJ0453+6224 & $55.0^f$  & 15340 & 260 &  1300 & 240 \\
ZTFJ0003+1404 &  $55.5^b$  & 12370 & 1255	& &\\
SDSS J1552+3201 & $56.3^a$  & 19570 & 1440 & 2180 & 335 \\
ASASSN-21au  &  $58.4^g$  & 15320 & 450	 & &\\
SDSS J1137+4054 & $59.6^h$  & 12640 & 260 &  1900 & 140\\
ZTFJ1637+4917 &  $61.5^b$  & 13010 & 420 &  2680 & 440 \\
SDSS J1319+5915 & $65.6^a$  & 12320 & 210 &  2160 & 180 \\
SDSS J1505+0659 & $67.8^i$  & 11090 & 220 &  2570 & 200 \\
    \end{tabular}
    \caption {Temperatures obtained for the hot and cold components of the AM CVns analyzed in this study and as shown in Figure \ref{fig:fig1}. $^a$ \cite{2018ramsay}, $^b$ \cite{2022vR}, $^c$ \cite{2021vR}, $^d$ \cite{2019RS}, $^e$ \cite{2020RS}, 
    $^f$ \cite{2023Rodriguez}, $^g$ \cite{2022RS}, $^h$ \cite{2021RS}, $^i$ \cite{2020Green}.}
    \label{tab:data_figure_1}
\end{table}
\bibliography{sample631}{}
\bibliographystyle{aasjournal}

\end{document}